\documentclass{elsart}

\usepackage{epsfig,psfrag}
\usepackage{amsmath,amssymb,bm}

\newcommand{\ket}[1]{\left|#1\right\rangle}
\newcommand{\erw}[1]{\left\langle#1\right\rangle}
\def\Xint#1{\mathchoice
   {\XXint\displaystyle\textstyle{#1}}%
   {\XXint\textstyle\scriptstyle{#1}}%
   {\XXint\scriptstyle\scriptscriptstyle{#1}}%
   {\XXint\scriptscriptstyle\scriptscriptstyle{#1}}%
   \!\int}
\def\XXint#1#2#3{{\setbox0=\hbox{$#1{#2#3}{\int}$}
    \vcenter{\hbox{$#2#3$}}\kern-.5\wd0}}

\def\dashint{\Xint-}

\begin{document}
\begin{frontmatter}

\title{Dynamics of a large spin with weak dissipation\thanksref{dedication}}
\thanks[dedication]{Dedicated to the 60th birthday of Ulrich Weiss}

\author[Hamburg]{T. Vorrath},
\author[Manchester]{T. Brandes}, and 
\author[Hamburg]{B. Kramer}
\address[Hamburg]{Universit\"at Hamburg, I.~Institut f\"ur Theoretische Physik,
                  Jungiusstr.~9, 20355~Hamburg, Germany}
\address[Manchester]{Department of Physics, University of Manchester,
                  UMIST, Manchester~M60~1QD, UK}

\begin{abstract}
We investigate the generalization of the spin-boson model to arbitrary spin
size. The Born-Markov approximation is employed to derive a master equation
in the regime of small coupling strengths to the environment. For spin one
half, the master equation transforms into a set of Bloch equations, the
solution of which is in good agreement with results of the spin-boson model
for weak ohmic dissipation. For larger spins, we find a superradiance-like
behavior known from the Dicke model. The influence of the nonresonant bosons
of the dissipative environment can lead to the formation of a beat pattern in 
the dynamics of the $z$-component of the spin. The beat frequency is 
approximately proportional to the cutoff $\omega_c$ of the spectral function.
\end{abstract}

\end{frontmatter}

\section{Introduction}
\label{sec_introduction}
The influence of a dissipative environment on the behavior of a two-state
system is often studied within the spin-boson model \cite{weiss}. 
This model applies to many 
systems from various fields of physics such as a magnetic flux in a SQUID,
electrons tunneling in chemical systems or double quantum dots, and 
two-level systems in glasses \cite{leggett}. Naturally, there are other
systems which are described by a spin greater than one half.
Nuclear spins constitute one example. The elements gallium and arsenic 
used in the majority of modern solid state experiments have a nuclear spin 
of 3/2. The nuclear relaxation process measured in NMR experiments 
\cite{stuttgart} is determined by the interaction with a dissipative 
environment, for instance a two-dimensional electron gas \cite{apel}.
Recent experiments have also been performed on molecular magnets. These are 
small clusters of a few atoms embedded into a crystal, which can be described
as large spins,
the most prominent examples of which (Mn${}_{12}$ and Fe${}_{8}$) are believed 
to have a total spin of 10 \cite{sessoli,wernsdorfer}.

Large spins also describe dissipation-induced collective effects of an 
ensemble of identical two-level systems. The coupling to the common dissipative
environment introduces an indirect interaction between the otherwise 
independent systems. A pseudo-spin -- formally the sum of all spin one halfs --
gives information about the degree of polarization of the ensemble.

In this work, we employ a master equation for the description of the 
large spin derived within the Born-Markov approximation.
The validity of the latter seems to be sometimes controversial 
in the literature, and our aim is to demonstrate that a careful derivation 
in the exact eigenstate basis of the coherent system indeed yields 
reliable results down to zero temperature for all parameter values, 
as long as the dissipative coupling (parameter $\alpha$) is small.

\section{Model Hamiltonian}
We take the examples of section~\ref{sec_introduction} as a motivation 
to study the generalization of the spin-boson model to larger spins, 
referred to as the \emph{large-spin model} in the following. 
The Hamiltonian is given by
\begin{equation}
\label{eq_hamiltonian}
H = \varepsilon \, J_z + 2 T_c \, J_x 
    + J_z \sum_q \gamma_q \, (a_q^{\dagger}+a_{-q})
    + \sum_q \omega_q \, a_q^{\dagger} a_q.
\end{equation}
The $J_i$ are components of a spin vector $\bm{J}$ of arbitrary size $J$.
For $J\!=\!1/2$, the model reduces to the spin-boson system. The bias is given
by $\varepsilon$ and $T_c$ accounts for tunneling between adjacent eigenstates
of $J_z$. 
The dissipative environment is modeled by a bath of harmonic oscillators
with creation operators $a_q^{\dagger}$ for a boson in mode $q$ and interaction
strength $\gamma_q$ to the spin. The collective character of the model is
due to the spin algebra which leads to non-constant tunnel rates
for transitions between the different states of the system.

The generalization of the spin-boson model to a dissipative multistate system 
has been studied by several authors \cite{schmid,guinea,fisher,weiss}.
There, a particle tunnels between different sites with a constant tunnel rate,
equivalent to a tight-binding model. 
Another possible generalization of the spin-boson model arises if also
excited states in a double-well potential are considered 
\cite{thorwart_2000,thorwart_2001}.
All these systems do not show any collective behavior by definition.

The Hamiltonian~(\ref{eq_hamiltonian}) also differs from the above cited 
generalizations of the spin-boson model with respect to the interaction term.
No counter term in the form of $J_z^2$ is included in 
equation~(\ref{eq_hamiltonian}). Such a term does not occur if e.~g. 
collective effects of an ensemble of two-level systems are described by 
the large spin. However, it is interesting to note in this context that a 
quadratic term, $J_z^2$, can appear in the opposite limit, the strong-coupling
regime, of the large-spin model. This is the case if a polaron transformation
is applied to the Hamiltonian~(\ref{eq_hamiltonian}). That term has 
considerable consequences on the dynamics of the large 
spin~\cite{vorrath_mb11}.

The Hamiltonian of a large spin with dissipation is closely related to the 
\emph{Dicke-model}, described by the Hamiltonian
\begin{equation}
\label{eq_dicke}
H_{\rm Dicke} = \omega_{\rm D} \, J_z + J_x \sum_q \gamma_q \, 
 (a_q^{\dagger}+a_{-q}) + \sum_q \omega_q \, a_q^{\dagger} a_q.
\end{equation}
The Dicke Hamiltonian is frequently applied in the field of quantum optics,
where it describes an ensemble of independent two-state atoms interacting via
the common radiation field. This model exhibits the effect of 
\emph{superradiance} as pointed out by Dicke in the original work in 1954,
Ref.~\cite{dicke}.
The decay of an ensemble of initially excited atoms is not exponential anymore,
as one would expect from independent atoms. Instead, the time of the decay
decreases inversely with the number of atoms and the maximum intensity of the
emitted radiation increases with the square of the number of atoms. 
The Dicke model does not include direct, electrostatic 
tunnel coupling terms. The 
coupling to the environment is offdiagonal in contrast to the spin-boson model
where the coupling is diagonal, i.e. to the $z$-component of the spin.
For zero bias, $\varepsilon\!=\!0$, the large-spin Hamiltonian reduces to the
Dicke model, though in a rotated frame of reference. Thus, the 
$x$-component of the large-spin model shows superradiant behavior in an
unbiased system. We will find that a similar statement applies for the 
$z$-component of the large spin in a biased system.

\section{Master Equation in Born-Markov Approximation}
We consider the regime of weak interactions between the large spin and 
the dissipative environment. The Born-Markov approximation is employed to 
derive a master equation for the density matrix of the spin. This 
method is perturbative in the system-reservoir coupling. In second
order, the master equation for the density matrix $\rho(t)$ of the spin reads
\cite{carmichael}
\begin{equation}
\label{eq_master}
\dot{\tilde{\rho}}(t) = - \int_0^t \! dt' \; 
  {\rm Tr}_{\rm Res} \Big\{  \big[ \tilde{V}(t), 
     \big[ \tilde{V}(t'), \tilde{\rho}(t) \otimes R_0 \big] \big] \Big\}.
\end{equation}
The interaction part of the Hamiltonian, Eq.~(\ref{eq_hamiltonian}), 
is denoted by $V$, the equilibrium density matrix of the environment by $R_0$,
and the tilde indicates the interaction picture.
The advantage of a master equation in this form is that only operators 
refering to the spin system enter. The degrees of freedom of the reservoir 
are already traced out with the understanding
that backaction effects on the bosonic bath can be neglected.

In (\ref{eq_master}), it is assumed 
that the spin is brought into contact
with the environment at time $t\!=\!0$. Hence, the initial density matrix of 
the system factorizes in a spin and a reservoir part. For finite times,
$t\!>\!0$, correlations between the subsystems evolve. These correlations
are neglected on the right hand side of equation~(\ref{eq_master}) in the 
second order Born approximation as they lead to effects of third or higher 
order in the coupling.
Moreover, the Markov approximation assumes that the kernel of the
integration in Eq.~(\ref{eq_master}) given by terms like
${\rm Tr}_{\rm Res}\{ \tilde{V}(t)\tilde{V}(t')R_0\}$ decays on a timescale
much shorter than the typical timescale of the spin density matrix 
$\tilde{\rho}(t)$. This is reasonable for weakly interacting systems since 
the dynamics of the spin in the interaction picture is solely caused by the 
coupling to the environment. Non-Markovian effects in the Born approximation 
for $J=1/2$ have been discussed recently by Loss and DiVincenzo \cite{LD03}.

Inserting $\tilde{V}(t)$ in the master 
equation~(\ref{eq_master}) and transforming back into Schr\"odin\-ger picture
yields the final form of the master equation for a large spin,
\begin{equation}
\label{eq_master-final}
\begin{split}
\dot{\rho}(t) =  \; &i\, \big[ \rho(t) , \varepsilon J_z + 2 T_c\, J_x \big] 
- \frac{1}{\Delta^2} \, (\varepsilon^2 \, \Gamma + 4 T_c^2 \, \Gamma_c) \,
        \big[ J_z, J_z \, \rho(t)  \big] \\[1mm]
&- \frac{2 T_c \, \varepsilon}{\Delta^2} \, (\Gamma - \Gamma_c) \,
        \big[ J_z, J_x \, \rho(t)  \big] 
+ \frac{2 T_c}{\Delta} \, \Gamma_s \,
        \big[ J_z, J_y \, \rho(t)  \big] \\[1mm]
&+ \frac{1}{\Delta^2} \, (\varepsilon^2 \, \Gamma^* + 4 T_c^2 \, \Gamma_c^*) \,
        \big[ J_z,  \rho(t) \, J_z \big] \\[1mm]
&+ \frac{2 T_c \, \varepsilon}{\Delta^2} \, (\Gamma^* - \Gamma_c^*) \,
        \big[ J_z,  \rho(t) \, J_x \big] 
- \frac{2 T_c}{\Delta} \, \Gamma_s^* \,
        \big[ J_z,  \rho(t) \, J_y \big].
\end{split}
\end{equation}
Here, the influence of the environment is expressed  by the rates
\begin{equation}
\label{eq_rates}
\begin{split}
\Gamma_c &= \frac{\pi}{2} \;\rho(\Delta) \,
             \coth\!\Big(\frac{\beta \Delta}{2}\Big)
 - \frac{i}{2} \; \dashint_0^{\infty} \! d\omega \; \rho(\omega) \;
 \Big( \frac{1}{\omega + \Delta} + \frac{1}{\omega - \Delta} \Big) ,\\[2mm]
\Gamma_s &= \frac{1}{2} \; \dashint_0^{\infty} \! d\omega \; \rho(\omega) \,
 \coth\!\Big(\frac{\beta \omega}{2}\Big) \; 
 \Big( \frac{1}{\omega + \Delta} - \frac{1}{\omega - \Delta} \Big)
 - i \, \frac{\pi}{2} \;\rho(\Delta),
\end{split}
\end{equation}
and $\Gamma\!=\!\Gamma_c(\Delta\!\to\!0)$, 
where $\Delta\!=\!\sqrt{4T_c^2\!+\!\varepsilon^2}$ is the level spacing of the
unperturbed large spin. In the following, we focus on the case of an ohmic 
dissipation when the spectral function $\rho(\omega)$ is linear with an 
exponential cutoff,
$\rho(\omega)\!=\!2 \alpha \,\omega \exp(-\omega/\omega_c)$.
Then, analytic expressions can be obtained for the rates except for the real
part of $\Gamma_s$ which remains to be calculated numerically at finite
temperatures.

\section{Relation to the Spin-Boson Model}
For the smallest possible value of the spin, $J\!=\!1/2$, the large-spin
Hamiltonian, Eq.~(\ref{eq_hamiltonian}), reduces to the spin-boson model. 
This enables us to compare the Born-Markov approximation with results for 
the spin-boson model in the literature. In particular, we employ the 
approximate solutions of the latter as given by Weiss \cite{weiss}.

It is a priori clear that the Born-Markov approximation being perturbative 
in the spin-environment coupling should only be applied in the regime of 
small couplings. Within the Born-Markov approximation, the spin is described
by the master equation~(\ref{eq_master-final}).
A peculiarity occurs for spin one half:
In that case, a closed set of equations for the expectation values of the
spin components follows from the master equation, comparable to 
Bloch equations,
\begin{equation}
\label{eq_spin-bloch}
\begin{split}
\dot{\erw{J_x}} &= 
 - \frac{1}{\Delta^2} \,
 \big( \varepsilon^2 \,{\rm Re}\{\Gamma\} + 4T_c^2\,{\rm Re} \{\Gamma_c\} \big)
   \erw{J_x} - \varepsilon \erw{J_y} \\[2mm]
 &\mspace{20mu} + \frac{2 T_c \, \varepsilon}{\Delta^2} \,
   \big( {\rm Re}\{\Gamma\}-{\rm Re}\{\Gamma_c\} \big) \erw{J_z} 
 + \frac{T_c}{\Delta} \, {\rm Im}\{\Gamma_s\} , \\[2mm]
\dot{\erw{J_y}} &= \varepsilon \erw{J_x} 
 - \frac{1}{\Delta^2} \,
 \big( \varepsilon^2 \,{\rm Re}\{\Gamma\} + 4T_c^2\,{\rm Re} \{\Gamma_c\} \big)
   \erw{J_y} \\[2mm]
 &\mspace{20mu}- \Big(2T_c + \frac{2 T_c}{\Delta} \, {\rm Re}\{\Gamma_s\} \Big)
   \erw{J_z}
 + \frac{T_c \, \varepsilon}{\Delta^2} \, 
   \big( {\rm Im}\{\Gamma\}-{\rm Im}\{\Gamma_c\}\big) , \\[2mm]
\dot{\erw{J_z}} &= 2 T_c \erw{J_y} .
\end{split}
\end{equation}
A similar set of Bloch equations was recently published by 
Hartmann et~al.~\cite{haenggi}. It is, however, not possible to apply these 
equations to larger spins without additional approximations. This was already
remarked in the original works by Bloch \cite{wangsness,bloch}.
The difficulty arises because products of spin operators cannot be replaced 
by a single spin operator for spins larger than one half. The 
corresponding equations do not form a closed set anymore. Hence, 
we can only transform the master equation into Bloch equations for spin 
one half. For larger spins we will have to come back to the master 
equation~(\ref{eq_master-final}).

The equilibrium values of the spin components follow from the Bloch 
equations~(\ref{eq_spin-bloch}). It is readily seen that 
$\erw{J_y}_{\infty}$ vanishes. The other two components become
\begin{equation}
\label{eq_equilibrium_bloch}
\begin{split}
\erw{J_z}_{\infty} &= \frac{\varepsilon \,
   {\rm Re}\{\varepsilon^2\, \Gamma + 4T_c^2 \,\Gamma_c\}\,
   {\rm Im}\{\Gamma\!-\!\Gamma_c\}
   + \varepsilon \Delta^3 \,{\rm Im}\{\Gamma_s\}}
{2\Delta \, {\rm Re}\{\varepsilon^2\, \Gamma + 4T_c^2 \,\Gamma_c\}
   (\Delta+{\rm Re}\{\Gamma_s\}) - 2\varepsilon^2 \Delta^2 
   {\rm Re}\{\Gamma\!-\!\Gamma_c\}} \,,\\[2mm]
\erw{J_x}_{\infty} &= 
 \frac{T_c \Delta^2 \,{\rm Im}\{\Gamma_s\} (\Delta + {\rm Re}\{\Gamma_s\})
   + \varepsilon^2 T_c \, {\rm Re}\{\Gamma\!-\!\Gamma_c\}\,
   {\rm Im}\{\Gamma\!-\!\Gamma_c\}}
 {\Delta \, {\rm Re}\{\varepsilon^2\, \Gamma + 4T_c^2 \,\Gamma_c\}\,
   (\Delta+{\rm Re}\{\Gamma_s\}) - \varepsilon^2 \Delta^2 \,
   {\rm Re}\{\Gamma\!-\!\Gamma_c\}} \,.
\end{split}
\end{equation}
In the limit of zero coupling, $\alpha \!\to\! 0$, we retrieve with 
Eq.~(\ref{eq_rates}) the thermodynamic expressions
\begin{equation}
\label{eq_thermodynamic}
\erw{J_z}_{\infty} = - \frac{\varepsilon}{2 \Delta} \,
                         \tanh \! \left( \frac{\beta \Delta}{2} \right),
\qquad
\erw{J_x}_{\infty} = - \frac{T_c}{\Delta} \,
                          \tanh \!\left( \frac{\beta \Delta}{2} \right).
\end{equation}
The dynamics of the spin components follows from the numerical integration
of the Bloch equations~(\ref{eq_spin-bloch}).

For a weak ohmic dissipation and $J=1/2$, Weiss gives an approximate solution for the
dynamics of the spin \cite{weiss}. Two temperature regimes have to be 
distinguished: At intermediate temperatures, the solution is obtained 
by the noninteracting-blip approximation (NIBA). At low temperatures, however,
the NIBA breaks down and the solution is derived by taking into account
interblip correlations. In the NIBA regime, the solution is given by 
equations~(21.132) and (21.134) in Ref.~\cite{weiss}.  The solution for low 
temperatures follows as equation~(21.172) and (21.173) in the same reference. 
Minor differences in the 
notation have to be remarked: The bias $\varepsilon$, the cutoff~$\omega_c$,
and the temperature $k_BT$ are identically defined in this article and in 
Ref.~\cite{weiss}. The coupling strength $\alpha$ corresponds to $K$ in 
Ref.~\cite{weiss}. The tunnel rate typically referred to as $\Delta$ in the 
spin-boson literature is expressed in our notation by $2 T_c$ (here, $\Delta$
is defined as the level spacing of the unperturbed spin).
Due to a different sign of the tunnel term
in the spin-boson Hamiltonian and the Hamiltonian for the large spin, 
Eq.~(\ref{eq_hamiltonian}), we have $\erw{\sigma_z}\!=\!2\erw{J_z}$ and
$\erw{\sigma_x}\!=\!-2\erw{J_x}$.

\begin{figure}
\centering
\psfrag{Jz}{\hspace*{-4mm} $\erw{J_z}$}
\psfrag{Jx}{\hspace*{-4mm} $\erw{J_x}$}
\psfrag{t}{\hspace*{-6mm} $t/ T_c^{-1}$}
\psfrag{jjz}{\footnotesize \hspace*{-2mm} $\erw{J_z}$}
\psfrag{jjx}{\footnotesize \hspace*{-2mm} $\erw{J_x}$}
\epsfig{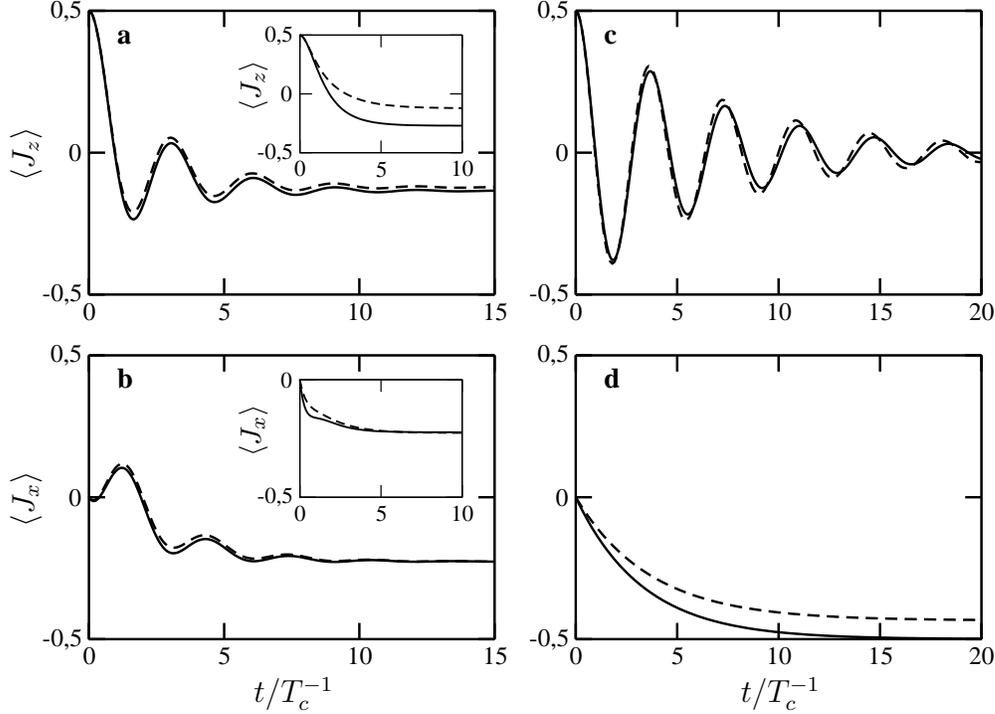}
\caption{\label{fig_spin} 
Dynamics of a spin 1/2 according to the Bloch equations~(\ref{eq_spin-bloch})
(solid line) and the approximate solutions by Weiss (dashed line); a and b:
$\varepsilon\!=\!T_c$, $\alpha\!=\!0.05$ (inset: $\alpha\!=\!0.2$),
$\omega_c\!=\!50T_c$, and $k_BT\!=\!2T_c$ 
(dashed line: Eqs.~(21.132) and (21.134) of Ref.~\cite{weiss}); 
c and d: $\varepsilon\!=\!0$, $\alpha\!=\!0.05$, $\omega_c\!=\!50T_c$, 
and $k_BT\!=\!0$ 
(dashed line: Eqs.~(21.172) and (21.173) of Ref.~\cite{weiss}).}
\end{figure}

Figures~\ref{fig_spin}a and \ref{fig_spin}b show the time evolution of $J_z$
and $J_x$ for weak ohmic dissipation, $\alpha\!=\!0.05$, at a finite 
temperature $k_BT\!=\!2T_c$.
The dynamics of $J_y$ follows as the time derivative 
of $J_z$, as can be seen from the Heisenberg equation of motion. For that 
temperature, the noninteracting-blip approximation applies and its solution
is in excellent agreement with the solution of the Bloch 
equations~(\ref{eq_spin-bloch}). Deviations between the two methods become
visible for a larger coupling strength, $\alpha\!=\!0.2$, as shown in the
inset of the same figure. However, this is not surprising as the Bloch
equations are perturbative in the coupling strength $\alpha$ and thus limited
to small couplings $\alpha$. Figure~\ref{fig_spin}c and \ref{fig_spin}d show 
the dynamics of an unbiased spin at zero temperature. In that regime, the 
NIBA is not valid anymore and the solution of the Bloch equations is compared 
to Eqs.~(21.172) and (21.173) of Ref.~\cite{weiss}. 
Again, both solutions are in good agreement.
For smaller couplings, e.~g. $\alpha\!=\!0.025$, the different solutions for
$J_z$ cannot be distinguished anymore (not shown).
The equilibrium value $\erw{J_x}_{\infty}$ of the Bloch equations appears 
too large as compared to the low temperature solution of the spin-boson model.
However, both equilibrium values approach the thermodynamic result, 
Eq.~(\ref{eq_thermodynamic}), in the limit of zero coupling and hence 
coincide in that limit.

We conclude that the Born-Markov approximation correctly describes the dynamics
of the spin-boson model for weak ohmic dissipation at {\em all} temperatures.
This is corroborated by recent results for the driven two-state 
system~\cite{haenggi}.
As we see no reason why the Born-Markov approximation should break down
for larger spins, we expect that the master equation~(\ref{eq_master-final}) 
gives a reliable description of a large spin with weak dissipation.

\section{Dicke Effect}
\begin{figure}
\centering
\psfrag{t}{\hspace*{-4mm}$t/ T_c^{-1}$} 
\psfrag{Jz}{\hspace*{-6mm} $\erw{J_z}/J$} 
\psfrag{J05}{\small $J=1/2$}
\psfrag{J2}{\small $J=2$}
\psfrag{J5}{\small $J=5$}
\psfrag{J10}{\small $J=10$}
\epsfig{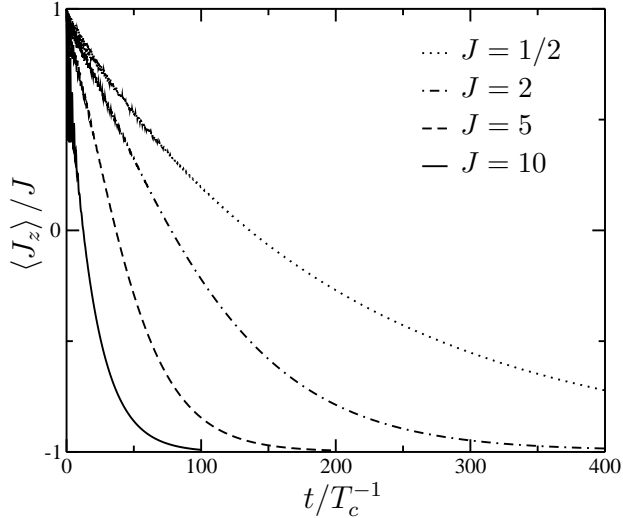}
\caption{ \label{fig_superradiance}
Decay of $J_z$ for different spin sizes
($\varepsilon\!=\!10 T_c$, $\alpha\!=\!0.005$, 
$\omega_c\!=\!50T_c$, $k_BT\!=\!T_c$).}
\end{figure}

It was pointed out in the introduction that the large-spin Hamiltonian,
Eq.~(\ref{eq_hamiltonian}), with zero bias can be mapped on the Dicke 
model, Eq.~(\ref{eq_dicke}), by rotation around the (spin) $y$-axis,
\begin{equation}
H_{\rm Dicke} = e^{i \pi/2 \, J_y} H \, e^{-i \pi/2 \, J_y} \,.
\end{equation}
The effect of superradiance characteristic for the 
Dicke model is therefore visible in the $x$-component of the large spin. This
becomes apparent if the initial values are chosen such that 
$\erw{J_x}_0\!=\!J$. Then, the effect results in an accelerated decay of $J_x$
with increasing spin size $J$. Typically, the main interest lies in the 
$z$-component of the spin and this is chosen maximum as initial value.
Consequently, the question arises if that component also decays in a 
superradiant fashion. This is indeed the case as will be shown in the 
following. We choose a large bias, $\varepsilon\!=\!10T_c$, to ensure the 
decay of $J_z$. The dynamics for different spin sizes $J$ follows from the
numerical solution of the master equation~(\ref{eq_master-final}).
The decay of the normalized $z$-component $\erw{J_z}/J$ is plotted in 
figure~\ref{fig_superradiance} for the spin sizes $J\!=\!1/2$, 2, 5, and 10.
It is clearly visible that the time in which the spin decays decreases with 
increasing spin size, similar to the Dicke superradiance.

Naturally, the exact dynamics of the large spin differs from the Dicke model
as the Hamiltonians are not identical. In contrast to the Dicke model, the
coupling to the environment is diagonal and a tunnel term exists in the 
unperturbed Hamiltonian. Yet, despite these differences, the basic results 
are the same. The reason is that superradiance is driven by the spin
algebra on which both models rely. 
This becomes obvious by considering the state dependent transition 
matrix elements \cite{dicke} which are maximal for the state
$\ket{J,M\!=\!0}$, corresponding to $\erw{J_z}\!=\!0$.
From the point of view of the spin-boson model it appears that the spin
shows collective effects, namely a superradiance like behavior,
once it is generalized to spins larger than one half.

\section{Quantum Beats}
\begin{figure}
\centering
\psfrag{t}{\hspace*{-4mm}$t/ T_c^{-1}$} 
\psfrag{g1}{\small $\alpha=0.0025$}
\psfrag{g2}{\small $\alpha=0.05$}
\psfrag{g3}{\small $\alpha=0.01$}
\psfrag{g4}{\small $\alpha=0.025$}
\epsfig{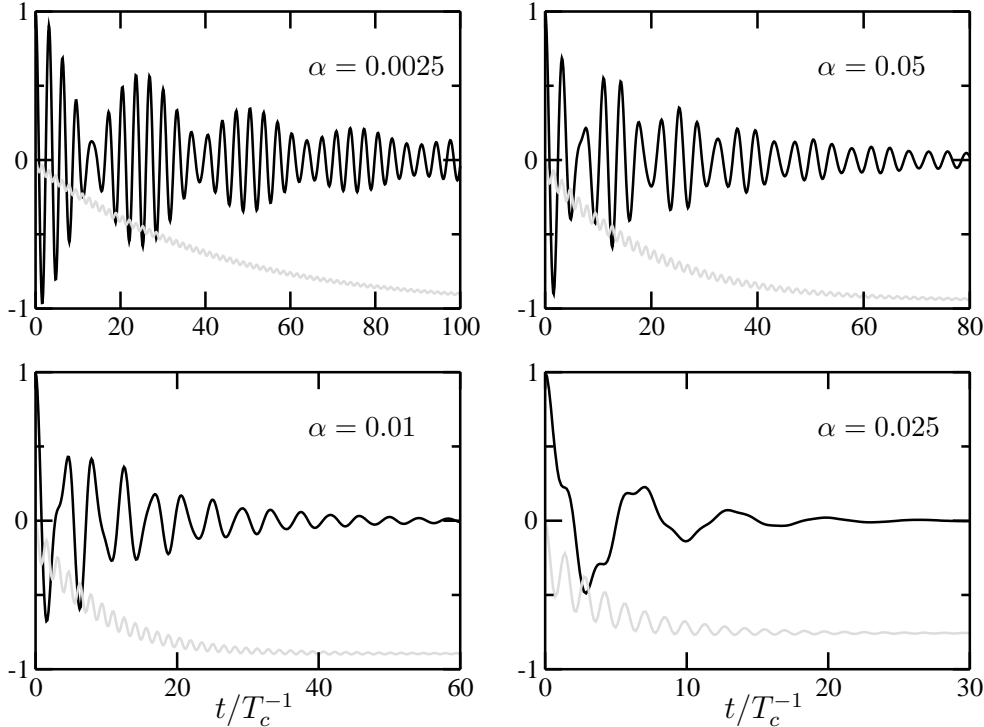}
\caption{ \label{fig_beats}
Time evolution of $J_z$ (black line) and $J_x$ (grey line) for different 
interaction strengths with the environment
($J\!=\!1$, $\varepsilon\!=\!0$, $\omega_c\!=\!50T_c$, and $k_BT\!=\!0$). }
\end{figure}

We shall return once again to the symmetric system, $\varepsilon\!=\!0$.
For a spin one half, the Bloch equations~(\ref{eq_spin-bloch}) predict an
exponential decay for $J_x$ and a damped oscillation for $J_z$. Already for
the next higher spin, $J\!=\!1$, new features appear in the dynamics of the
spin. Figure~\ref{fig_beats} shows the time evolution of the expectation
values of $J_z$ and $J_x$ for different interaction strengths. A clear 
beat pattern is visible on top of the damped oscillations of $J_z$. 
With increasing interaction with the environment, the beat pattern dissolves 
to a seemingly chaotic behavior. The decay of $J_x$ is superposed with 
oscillations. These new features have different origins. The beat pattern of
$J_z$ is caused by the nonresonant bosons of the environment. The oscillations
of $J_x$, on the other hand, are due to two-boson processes. The latter is 
best understood in a rotated frame, i.~e. in the Dicke model. There,
operator combinations $J_+ J_+$ and $J_- J_-$ appear in the master equation.
In standard approaches to superradiance~\cite{gross}, 
these are often disregarded by invoking  a \emph{secular approximation}. One can
show, however, that these terms lead to an additional oscillation the 
frequency of which is approximately given by $4T_c$, that is twice the level
spacing of the unbiased spin. We numerically find that the amplitude of these
oscillations is increased by the nonresonant bosons.

The beat pattern in the dynamics of $J_z$ is solely caused by nonresonant
bosons. They lead to different corrections to the eigenenergies of the spin,
lifting the equidistance of the spectrum. We consider a spin one with zero
bias at zero temperature, the same parameters as used in Fig.~\ref{fig_beats}.
The three eigenstates of the unperturbed spin labeled in the following by
$\ket{+}$, $\ket{0}$, and $\ket{-}$ have the eigenenergies $2 T_c$, 0, and
$-2T_c$, respectively. The expectation value of $J_z$ becomes 
$\erw{J_z}\!=\!\cos(2 T_c t)$. The nonresonant bosons of the environment
lead to corrections to these eigenenergies. In second order 
perturbation theory, we find
\begin{equation}
E_{\ket{\pm}}^{(2)} = - \frac{1}{2} \; \dashint_0^{\infty} \!d\omega \, 
\rho(\omega) \, \frac{1}{\omega\mp 2 T_c} \,, \qquad
E_{\ket{0}}^{(2)} = E_{\ket{+}}^{(2)} + E_{\ket{-}}^{(2)} \,.
\end{equation}
As a consequence, the three states are not equidistant anymore. The resonant
bosons, $\omega_q\!=\!2T_c$, are neglected in this calculation. They lead to
an imaginary part in the energies resulting in damping, which is not our 
concern at this stage. The next step is to determine the effect of the 
corrections on the dynamics of $J_z$. Neglecting the corrections to the
eigenstates themselves yields
\begin{equation}
\label{eq_beat}
\erw{J_z}_t = \cos(\omega_0 t) \cos(\omega_b t)
\end{equation}
with $\omega_0$ and $\omega_b$ defined as
\begin{equation}
\omega_0 = 2 T_c + \frac{1}{2} \,
           \big( E_{\ket{+}}^{(2)}-E_{\ket{-}}^{(2)} \big), \qquad 
\omega_b = - \frac{1}{2} \, E_{\ket{0}}^{(2)}.
\end{equation}
As a result, the oscillation frequency $\omega_0$
is slightly changed with respect to the unperturbed
case and we find indeed a beat pattern with frequency $\omega_b$. For an ohmic
dissipation, the beat frequency becomes
\begin{equation}
\omega_b = \alpha \, \omega_c + \alpha \, T_c \Big[ \,
  e^{2T_c/\omega_c} \, {\rm Ei}\Big(\frac{-2T_c}{\omega_c}\Big) 
  -  e^{-2T_c/\omega_c} \, {\rm Ei}\Big(\frac{2T_c}{\omega_c}\Big) \, \Big],
\end{equation}
where ${\rm Ei}(x)$ denotes the exponential integral.
The term in the brackets is of the order of $T_c/\omega_c$. Hence, 
for $\omega_c\gg T_c$, the beat
frequency is given in good approximation by $\omega_b\!=\!\alpha \omega_c$.
For weak interactions, the resulting dynamics is in excellent agreement with
the numerical solution of the master equation. Naturally, 
equation~(\ref{eq_beat}) does not include damping since dissipative effects 
of the resonant bosons are not considered in the derivation. 
The occurrence of a beat pattern is not restricted to the parameters chosen
in this example. Similar patterns are equally found for larger spins, 
$J\!>\!1$, and finite temperatures. The pattern dissolves at high temperatures
or high spins. Even for a finite bias, $\varepsilon\!>\!0$, such patterns
can be observed. 

\section{Conclusion}
The collective character of the model becomes visible for larger
spins where we found a superradiance-like decay of the $z$-component of the
spin, the time scale of which decreases with increasing spin size.
The beat pattern in the dynamics of $J_z$ for intermediate spins 
is caused by the nonresonant bosons of the environment. 
We conjecture that the beat pattern is a characteristic
property of an intermediate spin with weak dissipation. 

In our study of a large spin coupled to a dissipative environment, 
the theoretical description applied to real spins as well as to 
dissipation-induced collective effects in pseudo-spin systems such 
as ensembles or arrays of two-level systems.
We derived a master equation for arbitrary spin $J$  within the 
Born-Markov approximation. When comparing our results for $J\!=\!1/2$ 
and ohmic dissipation to standard spin-boson NIBA (and beyond) 
calculations~\cite{weiss}, we found good agreement 
at all temperatures and arbitrary parameter values, as long as the 
dissipative coupling is weak. 
We conclude that the master equation in Born-Markov approximation
is a reliable method to describe the physics in the weak coupling limit.


\begin{thebibliography}{00}
\bibitem{weiss} U.~Weiss,
  \emph{Quantum Dissipative Systems}
  (World Scientific, Singapore, 1999).
\bibitem{leggett} A.~J.~Leggett, S.~Chakravarty, A.~T.~Dorsey,
  M.~P.~A.~Fisher, A.~Garg, and W.~Zwerger,
  Rev. Mod. Phys. \textbf{59}, 1 (1987).
\bibitem{stuttgart}
  S. Kronm\"uller, W. Dietsche, K. v. Klitzing, G. Denninger, W. Wegscheider, 
  and M. Bichler, Phys. Rev. Lett. {\bf 82}, 4070 (1999); 
  J. H. Smet, R. A. Deutschmann, F. Ertl, W. Wegscheider, G. Abstreiter, 
  and K. v. Klitzing, Nature {\bf 415}, 281 (2002).
\bibitem{apel} W.~Apel and Yu.~A.~Bychkov, 
  Phys. Rev. Lett. {\bf 82}, 3324 (1999);
  Phys. Rev. B~\textbf{63}, 224405 (2001).
\bibitem{sessoli} R.~Sessoli, D.~Gatteschi, A.~Caneschi, and M.~A.~Novak,
  Nature \textbf{365}, 141 (1993).
\bibitem{wernsdorfer} W.~Wernsdorfer and R.~Sessoli,
  Science \textbf{284}, 133 (1999).
\bibitem{schmid} A.~Schmid,
  Phys. Rev. Lett. \textbf{51}, 1506 (1983).
\bibitem{guinea} F.~Guinea, V.~Hakim, and A.~Muramatsu,
  Phys. Rev. Lett. \textbf{54}, 263 (1985).
\bibitem{fisher} M.~P.~A.~Fisher and W.~Zwerger,
  Phys. Rev. B \textbf{32}, 6190 (1985).
\bibitem{thorwart_2000} M.~Thorwart, M.~Grifoni, and P.~H\"anggi,
  Phys. Rev. Lett \textbf{85}, 860 (2000).
\bibitem{thorwart_2001} M.~Thorwart, M.~Grifoni, and P.~H\"anggi,
  Ann. Phys. \textbf{293}, 15 (2001).
\bibitem{vorrath_mb11} T.~Vorrath, T.~Brandes, and B.~Kramer, 
  in \emph{Recent Progress in Many-Body Theories}, edited by R.~Bishop, 
  T.~Brandes, K.~Gernoth, N.~Walet, and Y.~Xian 
  (World Scientific, Singapore, 2002), p. 487.
\bibitem{dicke} R.~H.~Dicke,
  Phys. Rev. \textbf{93}, 99 (1954).
\bibitem{carmichael} H.~Carmichael,
  \emph{An Open Systems Approach to Quantum Optics}
  (Springer-Verlag, Berlin, 1993).
\bibitem{LD03}
  D.~Loss and D.~P.~DiVincenzo, cond-mat/0304118.
\bibitem{haenggi} L.~Hartmann, I.~Goychuk, M.~Grifoni, and P.~H\"anggi,
  Phys. Rev. E \textbf{61}, R4687 (2000).
\bibitem{wangsness} R.~K.~Wangsness and F.~Bloch, 
  Phys. Rev. \textbf{89}, 728 (1953).
\bibitem{bloch} F.~Bloch,
  Phys. Rev. \textbf{105}, 1206 (1957).
\bibitem{gross} M.~Gross and S.~Haroche,
  Phys. Rep. \textbf{93}, 301 (1982).
\end{thebibliography}
\end{document}